\documentclass[floatfix,twocolumn,aps,prl,showpacs,superscriptaddress]{revtex4}
\usepackage{amssymb}
\usepackage{graphicx}
\usepackage{dcolumn}
\usepackage{bm}

\begin{document}

\title{Thermal conductivity of quasi-one-dimensional 
antiferromagnetic spin-chain materials}
\author{A. V. Rozhkov}
\affiliation{Department of Physics, University of California, Irvine,
  California 92697} 
\author{A. L. Chernyshev}
\affiliation{Department of Physics, University of California, Irvine,
  California 92697} 
\date{\today }

\begin{abstract}
We study heat transport in quasi-one-dimensional  spin-chain
systems
by considering the model
of one-dimensional bosonic spin excitations interacting with
three-dimensional phonons and impurities 
in the limit of weak spin-lattice coupling and
fast spin excitations. A combined effect of the phonon and impurity
scatterings yields the following spin-boson thermal conductivity
behavior: $\kappa_s\propto T^2$ at low,
$\kappa_s\propto T^{-1}$ at intermediate, and 
$\kappa_s=const$ at higher temperatures.
Our results agree well
with the existing experimental data for 
Sr$_2$CuO$_3$.
We predict an unusual dependence on the impurity concentration for a
number of observables and propose further experiments.
\end{abstract}
\maketitle

{\it Introduction.}\ \ 
Recent experiments in one-dimensional (1D) quantum magnets
have revealed remarkably strong thermal transport anomalies 
associated with the low-dimensional spin degrees of freedom. 
In particular, an anisotropic thermal conductivity, comparable in
magnitude to that of metallic systems, 
was observed in quasi-1D spin-$\frac{1}{2}$ chain and ladder compounds
with a large part of the heat current attributed to magnetic
excitations.\cite{Sologubenko,Sologubenko1}
Only recently, however, a considerable progress in the theoretical
understanding of the thermal transport properties of 1D quantum spin
systems has been made.\cite{theories}

In this Letter we consider the heat transport in
spin-chain systems within the Boltzmann equation framework.
We develop a microscopic approach to the problem of 1D spin-boson excitations
interacting with the 3D phonon environment and impurities. An experimental
 realization of such a system is Sr$_2$CuO$_3$.\cite{Sologubenko}

Generally, in order to study transport problems one needs to identify
a mechanism 
responsible for the momentum relaxation. Conventional examples are the
Umklapp and 
impurity scatterings. Since the characteristic ``bandwidth'' for the
spin excitations 
in the real 1D spin chains materials of interest is very large ($J\sim
2000$K) in comparison with the experimental temperature range, the
Umklapp scattering  
of spin excitations on themselves is strongly suppressed (as $\sim e^{-J/T}$).
Instead, we propose that the leading relaxation
mechanism at not too low temperatures is the two-stage,
bottle-neck process of {\it (i)} transferring the momentum from spins
to phonons, and {\it (ii)} the subsequent dissipation of the
momentum via a phonon Umklapp process. 
The central idea is that the stage {\it (i)} is the bottle-neck
because the spin-lattice  
coupling is very weak. Thus, the excess momentum in the spin subsystem
waits the longest time to get transfered to phonons, but once it
is transferred it relaxes quickly. 
If this is the case we have to find only the spin-phonon
relaxation time $\tau_{sp}$ that now 
plays the role of transport relaxation time for the spin excitations.
However, as we demonstrate below, the spin-phonon scattering
mechanism alone becomes too weak for the low-lying excitations
which leads to an infinite spin thermal conductivity $\kappa_s$.
On the other hand, the impurity scattering is very strong in 1D at low
energies. 
This scattering mechanism renders $\kappa_s$ finite for any realistic system 
and allows us to provide a detailed description of the thermal transport.

{\it Model.} \ \ 
The Hamiltonian of a single Heisenberg antiferromagnetic spin-$\frac12$ 
chain is:
\begin{eqnarray}
\label{H0}
H_{\rm chain} = J \sum_i {\bf S}_i \cdot {\bf S}_{i+1}.
\end{eqnarray}
The interchain coupling is very small in Sr$_2$CuO$_3$,\cite{Sologubenko} 
$J^\prime /J\sim 10^{-5}$, and is neglected here.
After the Jordan-Wigner transformation followed by
bosonization Eq. (\ref{H0}) 
can be expressed as the Hamiltonian of free bosonic field
$\Phi$, which we refer to as the spin-boson field:
\begin{eqnarray}
H_0 = \frac{v}{2} \int dx \left( \Pi^2 + \left( \partial_x
\Phi \right)^2 \right)= v\sum_k|k| b_k^\dag b_k^{\phantom{\dag}},
\end{eqnarray}
where $\Pi(x)$ is a conjugate field momentum, and velocity $v$ is given
by $ v \propto Ja, $ $a$ is the chain lattice constant,\cite{book} and
$b, b^\dagger$  
are the spin-boson operators.

Since the superexchange $J$ is a function of the inter-site distance 
the lattice vibrations can modify it. This
will couple phonons to the spin-boson Hamiltonian density. Such an
interaction can be 
written in terms of the phonon operators $a, a^\dagger$ and the spin-boson
operators as follows:
\begin{eqnarray}
&&H_{\rm sp} = -\frac{g_{sp}}{\sqrt{N}} \sum_{kk'{\bf P}\ell}
V_{\ell}({\bf P},k,k')
\left( a^\dagger_{{\bf P}\ell} b^\dagger_k b^{\vphantom{\dagger}}_{k'}
 + {\rm h.c.} \right)\ ,\label{Hint}\\
&& \ \ V_{\ell}({\bf P},k,k')=
\frac{P_\| k k'}{\sqrt{ 8m_i \omega_{{\bf P}\ell}kk'}} 
\left({\bm\xi}_{{\bf P}\ell}\right)_x\delta_{P_\|,k'-k}\ ,\label{Vint}
\end{eqnarray}
where $g_{sp}$ is a spin-phonon coupling constant, the index $\ell$
enumerates three  
phonon polarizations,
$\omega_{{\bf P}\ell} = c_\ell |{\bf P}|$ is the phonon dispersion,
${\bm \xi}$ is the polarization vector of a phonon, $m_i$ is the mass of the
unit cell, $N$ is the number of unit cells in the sample, $k$ and $k'$ are
the 1D momenta of spin bosons, and ${\bf P}$ is the 3D momentum of a phonon.
Note that only a component of the total momentum along the chain is
conserved, which is explicitly given by the $\delta$-function. Sound
velocities $c_\ell$ are much smaller than the spin-boson velocity $v$ for
the real spin-chain materials: $c_\ell \ll v$. Eq. (\ref{Hint})
describes the  processes in 
which spin boson either emits or absorbs a phonon.
The projection of the phonon polarization vector on the direction of
the chain ($x$-axis) 
$( {\bm\xi}_{{\bf P}\ell})_x$ is equal to: $|P_\| |/|{\bf P}|$ (longitudinal), 
and $\sqrt{1-(P_\|/{\bf P})^2}$ or $0$ (transverse), where two answers
for the transverse  
phonons correspond to the polarization in the plane given by $\hat x$
and ${\bf P}$ 
and normal to such a plane, respectively. The latter phonons do not
couple to  spins and are not discussed any further. 

The impurity coupling is the final component of the model. Impurities
lead to a  
local variation of the superexchange $J$, that leads to the scattering
of magnetic excitations. 
The most important (relevant in the renormalization group (RG) sense)
part of the impurity 
Hamiltonian for an impurity located at $x_{0}$ can be written as:
\begin{eqnarray}
H_{{\rm imp} }& = &\frac{\delta J_{{\rm imp} }}{\pi }
\cos \left( 2k_{\rm F} x_0 + \sqrt{4\pi{\cal K}} \Phi(x_0 )
\right),\label{Himp}
\end{eqnarray}
where the Luttinger-liquid parameter ${\cal K} = 1/2$.

{\it Phonon mechanism of relaxation.}
To study the relaxation of 1D spin excitations 
coupled to the bath of 3D phonons we solve the stationary Boltzmann equation:
$v \partial_x f_k = -S_k[f]$,
where $f_k$ is the spin-boson distribution function and $S_k[f]$ is the
collision integral, given by:
\begin{eqnarray}
\label{CE}
S_k[f]= \int_{k'}
\bigg[ W_{kk'}f_k (f_{k'} + 1)- W_{k'k}f_{k'}(f_k +
  1)\bigg]\ , 
\end{eqnarray}
where $\int_{k'}$ stands for $\int dk'/2\pi$ and
$W_{kk'}$ is the total probability of the spin excitation 
to be scattered from the state $k$ to the state $k'$.
For the processes of scattering due to 
phonons such probabilities are given by:
\begin{eqnarray}
\label{Wkk}
W_{kk'}=\sum_\ell \int_{{\bf P}}\bigg( w^\ell_{kk'{\bf P}} 
(n^0_{{\bf P}\ell} + 1)^{\vphantom{\ell}} +
w^\ell_{k'k{\bf P}} n^0_{{\bf P}\ell}\bigg)\ ,
\end{eqnarray}
where $\int_{{\bf P}}$ denotes $(1/2\pi)^3\int  d^3 {\bf P}$.
The probability $W_{k'k}\equiv W_{kk'}[k\leftrightarrow k']$. We
assume that for all relevant temperatures the phonon relaxation time is much 
shorter than that of the spin
bosons.  Consequently, one can use the equilibrium form for the 
phonon distribution function $n_{{\bf P}\ell}=n^0_{{\bf P}\ell} =
1/(e^{\omega_{{\bf P}\ell}/T} - 1 )$.
The ``elementary'' scattering probabilities $w^\ell_{kk'{\bf P}}$
of the spin boson due to emission or absorption of the
phonon with the momentum ${\bf P}$ and polarization $\ell$
are determined from Eqs. (\ref{Hint}), (\ref{Vint}):
\begin{eqnarray}
w^\ell_{kk'{\bf P}} &=& \frac{g_{sp}^2V_0}{8m_i}\cdot
\frac{P_\|^2|kk'|}{\omega_{{\bf P}\ell}} ({\bm \xi}_\ell)^2_x
\label{w}\\
&&\times \delta(k'+P_\|-k) \delta (\omega_{k'} + \omega_{{\bf P}l}-\omega_k),
\nonumber
\end{eqnarray}
where $V_0$ is the elementary cells volume, $\omega_k = v|k|$ is the spin-boson
energy.
Note that the total energy and the projection of the 
momentum along the chain are conserved, which is
enforced by the $\delta$-functions in Eq. (\ref{w}).

For a small temperature gradient $ \partial_x T \ne 0$
one can introduce a
usual ansatz for the distribution function $f_k$ to
linearize the Boltzmann equation:
$f_k = f_{k}^0 + f_{k}^1$,
where $f_{k}^1$ is a non-equilibrium correction to the
equilibrium Bose distribution function $f_{k}^0$. Function $f_k^1$ should be
 small and proportional to the temperature gradient.
Thus, the Boltzmann equation for the spin bosons to the first order in
$\partial_x T$ can be written as:
\begin{eqnarray}
\label{BE_lin}
\frac{v|k|}{T} \left(\partial_x T \right) \frac{\partial f_k^0}{\partial k}
= S_k,
\end{eqnarray}
where $S_k$ is the linearized collision integral given by 
$S_{k}
=\sum_\ell\left(S^{(1)}_{k\ell}+S^{(2)}_{k\ell}\right)$ with 
\begin{eqnarray}
\label{S1}
S^{(1)}_{k\ell} =
\int_{k'{\bf P}} w^\ell_{kk'{\bf P}} \bigg[  P^{(1)}_{k,{\bf P}}\  f_{k}^1   
+  P^{(2)}_{k,{\bf P}} \ f_{k'}^1 \bigg] \ ,
\end{eqnarray}
where $P^{(1)}_{k',{\bf P}}=\left( n_{{\bf P}\ell}^0
+ f_{k'}^0 + 1\right)$, $P^{(2)}_{k,{\bf P}}=\left(f_{k}^0-n_{{\bf
    P}\ell}^0 \right)$ 
and $S^{(2)}_{k\ell}\equiv -S^{(1)}_{k\ell} [k\leftrightarrow k']$.
The integral $S^{(1)}$ accounts for two types of collision events:
{\it (i)}  spin boson with the momentum $k$ emits a phonon with the momentum
${\bf P}$ and scatters into the state with the momentum $k'$; {\it (ii)} 
process inverse to
{\it (i)}. Likewise, $S^{(2)}$ describes absorption of a phonon by the spin
boson with the momentum $k$ and the corresponding inverse process.

Once $f^1_k$ is found the
spin-boson  thermal current density $J_E^s$ can be written as:
$J_E^s = \int_k v^2 k f^1_k = -\kappa_s\partial_x T$,
where the coefficient $\kappa_s$ is the spin-boson thermal
conductivity. 
Note that the use of Boltzmann equation implies the semiclassical
transport, condition for which can be shown to be satisfied for the
real 1D spin system. 

{\it Kinematic considerations.}\ \ 
To solve the Boltzmann equation one needs to evaluate
integrals in Eq. (\ref{S1}). Part of
this task can be accomplished without any approximations as the 
integral over the phonon momentum ${\bf P}$ in $S^{(1)}$ and $S^{(2)}$ 
can be calculated explicitly.
The remaining integration over $k'$ is restricted to some finite
intervals due to  
the conservation of the energy and momentum.\cite{us_big}

Further, there is a following simplification. As we noted above,
without impurities  
the heat conductivity will diverge at small $|k|$. 
Thus, the biggest contribution to $\kappa_s$ comes from the
low-energy spin bosons: $v|k| \ll \min(T,\Theta_D)$ and we need to
consider only them. 

{\it Relaxation time.}\ \ 
One can write the collision integral in Eq. (\ref{S1}) as a
sum of two terms:
$S_{k} = {f_{k}^1}/{\tau_{sp}(k)} + \delta S_{k}$,
where the first term has the usual relaxation time form, while the
second one does not. However, the second term can be demonstrated to be
small for small spin-boson energy $\delta S_k/S_k\propto
(vk/T)^2$. 

Finally, with the help of this approximation, the collision integral
can be written in the relaxation time form $S_k \approx
f^1_k/\tau_{\rm sp}(k)$, where $\tau^{-1}_{\rm sp}(k)$ plays the role of the
transport relaxation rate and is given by:\cite{us_big}
\begin{eqnarray}
\label{tau_tr00}
\tau_{\rm sp}(k)^{-1} =
\frac{{{\cal A}}k^2 T^3}{v^3}\Gamma(\Theta_D/T)\ , 
\end{eqnarray}
where $\Gamma(z)=\frac{I_1(z)}{I_1(\infty)}$, and $I_1(z)=\int_0^z
\frac{x^3 dx}{2\sinh^2 x/2}$. 
Here the phenomenological constant ${\cal A}$ is of the
order of $(V_0 / m_i) (g_{sp} / c)^2$.
By evaluating the asymptotic properties of $\Gamma(z)$ one can easily
verify that: 
\begin{eqnarray}
\tau_{\rm sp}(k)^{-1} = \cases{{{\cal A} T^3 k^2}/{v^3}  & for
$T < \widetilde \Theta_D$,\cr
{{\cal A}{\widetilde\Theta_D}^2 T
 k^2}/{v^3} \ & for $T > \widetilde \Theta_D$,\cr}
\label{tau_tr0_a}
\end{eqnarray}
where ${\widetilde\Theta_D}=\Theta_D/\sqrt{I_1(\infty)}\approx
\Theta_D/4$ plays the role of a crossover temperature.
The relaxation rate for the spin boson,
Eq. (\ref{tau_tr0_a}), is our main result for the spin-phonon scattering.
We note that the $k$- and $T$-dependence of the bottle-neck relaxation
rate will be the same for other low-D spin systems with large $J$
where momentum is conserved in the dimensions fewer than D=3.

It is interesting to note that since the spin bosons are fast ($c \ll v$), the
energy and momentum conservation dictates that the majority of the
phonons which interact with the spin subsystem must have their
momentum almost normal to the chain direction. 
Indeed, the phonon energy is 
$\omega_{\bf P} = cP = v||k|-|k'||$, where $k$ and $k'$ are the
1D spin-boson momenta, while the momentum conservation along the chain
gives $|P_\|| = |k - k'|$. Therefore, for a typical phonon: 
$P \gg |P_\||$, except
for the case of almost elastic backward scattering ($k' \approx
-k$) which does not contribute substantially to the transport
relaxation time. Consequently, we determine
${\bm \xi}_{xl} = {\cal O} (c/v)$ and 
${\bm \xi}_{xt} = {\cal O}(1)$. Thus, the most effective spin-phonon 
scattering is due to the transverse phonons.

With the collision integral in the relaxation time form 
the Boltzmann equation is trivially solved:
\begin{eqnarray}
f^1_k = \frac{v|k|\tau_{\rm sp}(k)}{T} \left(\partial_x T \right)
\frac{\partial f_k^0}{\partial k} \approx -\frac{\tau_{\rm sp}(k)}{k}
(\partial_x T)\ , \label{f1_ph}
\end{eqnarray}
and the thermal conductivity is calculated:
\begin{eqnarray}
\kappa_s = \int v^2 \tau_{\rm sp}(k) \frac{dk}{2\pi}=  
\frac{v^5}{2\pi {\cal A} T^3} \int_0^{T/v} \frac{dk}{k^2}\ .
\label{k_diver}
\end{eqnarray}
This expression diverges at small $k$ giving rise to an infinite thermal
conductivity.  It occurs because the scattering of spin
bosons on phonons is not sufficiently strong for $k \rightarrow 0$ to
ensure the convergence of the low-energy contribution to the integral. 

{\it Impurities.} \ \
In order to remove the divergence of (\ref{k_diver}) we need to introduce
yet another scattering mechanism for the low-$k$ spin bosons. In a 1D system
impurities scatter low-energy excitations very effectively.
Therefore, disorder in the magnetic coupling $J$ will regularize the
infrared divergence in Eq. (\ref{k_diver}).
Since the spin-impurity interaction in Eq. (\ref{Himp}) 
is not a low-order polynomial in $\Phi$, the Boltzmann transport theory 
cannot be directly applied to the impurity scattering.
Instead, we will evaluate the spin boson life-time
$\tau_{\rm imp}$ using the
Green's function perturbative expansion in powers of $\delta J_{\rm imp}$.

There are two issues we must clarify before proceeding with this approach.
First, for the perturbation theory in powers of $\delta J_{\rm imp}$ to be
valid the temperature must be bigger than the Kane-Fisher 
temperature:\cite{KaneFisher} $T \gg T_{KF}$. For our system with
${\cal K} = 1/2$ the 
Kane-Fisher temperature is 
$T_{\rm KF} = { \delta J_{\rm imp}^2}/{J}$.
Second, it is important to note that, generally, the life-time of an
excitation  is not equivalent to the transport relaxation time. 
While the only processes which
violate the conservation of the momentum contribute to the latter, any kind of
scattering shortens the former. However, since Eq. (\ref{Himp})
does not conserve momentum and since in any typical
scattering event the spin-boson momentum changes
drastically, there is no distinction between these two time scales in
our case. This justifies the use of our approach. 

The second-order correction
to the spin-boson Green's function  from the impurity scattering is:
\begin{eqnarray}
\delta{\cal D}^{(2)}_{k}(\tau)=
\int_{\tau^\prime\tau^{\prime\prime}}
\left< B^\dagger_k(\tau)
\overline{H_{\rm imp} (\tau^\prime) H_{\rm imp} (\tau^{\prime\prime})}
B^{\vphantom{\dagger}}_{k'}(0)
\right> ,
\nonumber
\end{eqnarray}
where $B^\dagger_k(\tau)=b^\dagger_k(\tau) +
b^{\vphantom{\dagger}}_{-k}(\tau)$ and
the line over the Hamiltonians stands for the disorder averaging.
The right hand side of this equation is proportional to the Matsubara
self-energy. After analytical continuation we find the
retarded self-energy:\cite{us_big}
\begin{eqnarray}
\label{Sigma1}
\Sigma^R_{k,\omega} \approx
- i\ \frac{n \ \delta J_{\rm imp}^2 \omega}{a\pi^2 J |k|T} \ ,
\end{eqnarray}
with $n$ being the dimensionless impurity concentration. 
Once the self-energy is found one can obtain the dressed
Green's function:
${D}_{k,\omega} = {2\omega_k}/
({{\omega_k^2 - \omega^2} - 2 \omega_k \Sigma_{k,\omega}^R})$.
By finding the poles of this expression the damping of
the spin bosons is determined as given by:
\begin{eqnarray}
\label{tau_imp}
\tau_{\rm imp}^{-1}=  \frac{\Delta^2}{T},
\end{eqnarray}
where
$\Delta^2 \propto n \delta J_{\rm imp}^2$.\cite{Affleck}
Unlike in the spin-phonon case the impurity relaxation rate
is independent of $k$ and thus will remove the divergence of the thermal
conductivity.

{\it Thermal conductivity.}
Finally, having at hands the transport relaxation rates for the
spin boson due to phonon and impurity scattering, Eqs. (\ref{tau_tr00}) and (\ref{tau_imp}), 
we can calculate the thermal conductivity of
the spin chains. 
The total relaxation time can be found according to the
Matthiessen's rule:
${\tau_{\rm tot}}^{-1} = {\tau_{\rm imp}}^{-1} +
{\tau_{\rm sp}}^{-1}$.
Then the expression for the thermal conductivity (\ref{k_diver})
should be modified to:
\begin{eqnarray}
\label{kappa2}
\kappa_s (T) = \frac{v T^2}{\pi\Delta^2}
\int_0^{J/T} 
\frac{x^2 \ dx}{4\sinh^2(x/2)}\ \frac{\alpha(T)}{x^2 + \alpha(T)}\ ,
\end{eqnarray}
with
$\alpha(T)={v^5 \Delta^2}/{{\cal A} T^6\Gamma({\Theta_D}/{T})}$.
Since the temperature is always much smaller than $J$ one can safely
replace the upper limit in Eq. (\ref{kappa2}) by infinity.

Eq. (\ref{kappa2}) specifies a function with a single maximum at $T = T_m
\propto (n \delta J^2_{\rm imp}/{\cal A} J)^{1/6}$,
the maximum value of $\kappa_s$ is $\kappa_s^{max} = \kappa_s (T_m) \propto
v^6/ {\cal A} T_m^4$. The function $\kappa_s(T)$ vanishes as $T^2$ at
$T \ll T_m$ and saturates at $T \gg T_m$, approaching $\kappa_s^\infty \propto
\kappa_s^{max} T_m/\Theta_D$:
\begin{eqnarray}
\label{kappa_q1}
\frac{\kappa_s(T)}{\kappa_s^{max}}\approx\cases{ 
2.26\ {T^2}/{T_m^2} & for $T\ll T_m$ ,\cr
1.46\ {T_m}/{T} & for $T_m\ll T\ll {\widetilde\Theta_D}$ , \cr
5.55\  T_m/\Theta_D& for $T\gg {\widetilde\Theta_D}$ . \cr} 
\end{eqnarray}
At temperatures below $T_m$, $\kappa_s$ is controlled by
impurity scattering.\cite{mus'yu} At temperature $T>T_m$ the number of  
 higher energy spin bosons, which scatter mostly on phonons, increases.
However, the low-$k$ spin bosons are still
scattered mostly by impurities. Thus, at $T \gg T_m$ both mechanisms
contribute to $\kappa_s$.

{\it Comparison with experiments.}\ \ 
Using Eq. (\ref{kappa2}) we compare our theory with the experimental
data for Sr$_2$CuO$_3$.
It is convenient to measure the temperature in units of $T_m$ and the thermal
conductivity in units of $\kappa_s^{max}$. This guarantees that the location
of the maximum coincides with the maximum of the experimental
curve.
\begin{figure}[t]  
\includegraphics[width=8cm]{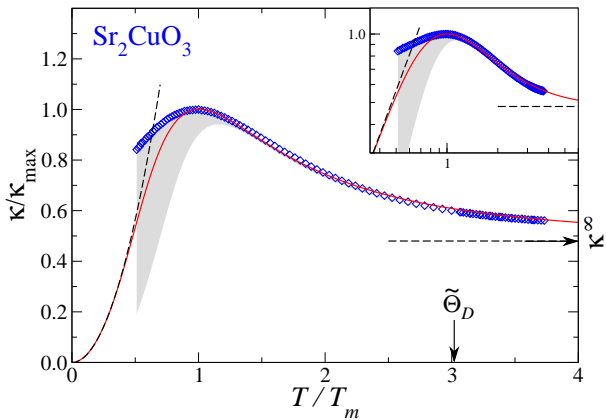}
\caption{Normalized spin thermal conductivity $\kappa_s$ 
vs reduced temperature $T/T_m$. Diamonds are the experimental data for
  Sr$_2$CuO$_3$, $T_m=79.4$K and
  $\kappa_s^{max}=36.7$Wm$^{-1}$K$^{-1}$, see
  Ref. \onlinecite{Sologubenko}.
Solid line is the result of this work,
  Eq. (\ref{kappa2}), for 
  $\Theta_D=11.6 T_m$. Arrows mark the saturation value
  $\kappa_s^\infty$ for the solid line at
  $T\gg {\widetilde \Theta_D}$ 
and the spin-phonon scattering crossover scale
  ${\widetilde \Theta_D}$, respectively. Asymptotic behavior at $T\ll
T_m$ ($\sim T^2$) 
and at $T\gg T_m$ ($\sim const$) are shown by the dashed lines. Inset:
same data on the log-log scale.}
\label{chain}
\end{figure} 
Once this rescaling is done there is only 
one parameter to adjust in our theory: $\Theta_D$.
By choosing
$\Theta_D = 11.6T_m$ we fit the experimental
data for $T > T_m$ almost ideally.
Shaded area at $T < T_m$
highlights the region where the phonon background
subtraction creates a large uncertainty in the experimental 
data.\cite{Sologubenko} With this in mind we conclude that Eq. (\ref{kappa2})
describes very well the whole range of available experimental data.
%

Our work predicts a remarkable behavior of the thermal conductivity at
higher temperatures: saturation at a constant value. 
A parallel can be drawn with metals, where $\kappa$ also saturates
at high temperatures.
%
However, in our case the excitations are not fermions at $E_F$, but 
the long-wavelength spin bosons. Because of that,
the manner in which impurity scattering cuts off such excitations at
low energies remains important even at high temperatures. 
Therefore, in the case of spin chains the
saturation of $\kappa_s$ is a result of a non-trivial combination of
several effects: temperature- and $k$-dependence of the
impurity and phonon relaxation rates and the 1D density of states of
the spin-bosons. We note that the saturated regime in $\kappa_s$ should 
persist up to $T\sim J$ above which bosonic 
description of excitations 
is not adequate.

{\it Predictions.}\ 
One of the most straightforward ways to verify our theory would be to
measure the thermal conductivity for higher temperatures and check if
the saturation really takes place as we predict.
Next, we see that the temperature $T_{m}$ at which $\kappa_s(T)$ reaches
its maximum scales as $n^{1/6}$.  
Also, the maximum value of the thermal
conductivity $\kappa_s^{max} = \kappa_s(T_{m})$ scales as $n^{-2/3}$,
an unusual and rather strong effect.
Finally,
the saturation value $\kappa_s^\infty$ has yet different concentration
dependence $n^{-1/2}$. 
If such a behavior is
observed in the
materials with the isotope substitution for either Cu or O ions
it will provide a strong support to our theory.

{\it Conclusions.}\ \ 
To conclude, we have developed a theory of anomalous heat transport in the 
spin-chain systems coupled to the 3D phonon
environment in the presence of weak disorder. 
We have calculated the thermal conductivity as a function of
temperature and have obtained that
the low-temperature transport is dominated by the impurity scattering while
the high temperature transport is determined  by both the  impurity
scattering and the spin-phonon collisions.
Our main results are in a very good quantitative agreement with the
available experimental data. 
In implementing our approach we have also obtained an insight into
various microscopic details of the problem.
This has allowed us to formulate several predictions and
suggest future experiments.

This work was supported by DOE under grant DE-FG02-04ER46174.


\end{document}